\documentclass[11pt,a4paper]{article}
\pdfoutput=1
\usepackage{jheppub}
\usepackage{graphicx}
\usepackage{amsmath}
\usepackage{amssymb}
\usepackage{hyperref}
\usepackage{natbib}
\usepackage{slashed}
\usepackage{xspace}
\usepackage[svgnames]{xcolor}
\usepackage[normalem]{ulem}
\usepackage[utf8]{inputenc}
\usepackage[T1]{fontenc}
\usepackage{xfrac}
\usepackage{comment}

\newcommand{\Td}{T'}

\newcommand{\eqsp}{\;\,}

\preprint{ULB-TH/22-11}

\title{The domain of a cannibal dark matter}
\author[a]{Marco Hufnagel}
\emailAdd{marco.hufnagel@ulb.be}
\author{and}
\author[a]{Michel~H.G.~Tytgat}
\emailAdd{michel.tytgat@ulb.be}
\affiliation[a]{Service de Physique Th\'eorique, Universit\'e Libre de Bruxelles, Boulevard du Triomphe, CP225, 1050 Brussels, Belgium}

\abstract{
We consider a scenario in which the dark matter is alone in a hidden sector and consists of a real scalar particle with a manifest or spontaneously broken $\mathbb{Z}_2$ symmetry, at a temperature which differs from the one of the visible sector, $\Td \neq T$. While similar models with general couplings have already been studied in the literature, the special case of a model with spontaneous symmetry breaking constitutes a non-trivial limit of these results, since it features vanishing tree-level amplitudes for the processes $k \rightarrow 2$ with $k>2$ at threshold, thus making the cross-section governing dark-matter freeze-out velocity suppressed. We carefully determine the thermally averaged dark-matter annihilation cross-section in this scenario, including the possible effects of one-loop corrections and Bose-Einstein statistics, while also reporting our results in the domain of thermal dark matter candidates, $\Td_\text{fo}/T$ vs.~$m_\text{DM}$ with $\Td_\text{fo}$ being the hidden-sector temperature at decoupling. We show that for fixed quartic coupling, due to entropy conservation, the thermal candidates lie along a curve $\Td/T \propto m_\text{DM}^{-\sfrac13 (1+\kappa)}$  with anomalous scaling $\kappa \sim \mathcal{O}(\Td_\text{fo}/m_\text{DM}) > 0$. Furthermore, we demonstrate that this scaling is valid for a broad class of cannibal DM scenarios, with or without symmetry breaking. In addition, we also discuss the relevant cosmological and astrophysical constraints.
}

\begin{document}
\maketitle

\section{Introduction}

The nature of dark matter (DM) is still shrouded in mystery. While a weakly interacting massive particle (WIMP) remains a simple and appealing possibility, the  lack of  experimental evidence for such states provides ever stronger motivation for also exploring alternatives. One promising direction is to consider hidden sector (HS) scenarios, which feature a DM candidate that significantly interacts with itself and potentially other dark-sector states, while only being -- at best -- feebly coupled to the Standard Model (SM), or more generally to the visible sector (VS).\footnote{Such scenarios are further motivated as a potential solution for some of the issues that collisionless DM faces on galactic and sub-galactic scales~\cite{Spergel:1999mh,Tulin:2017ara}.} Such a HS may be in thermal equilibrium with itself, but not with the VS, meaning that its temperature $\Td$ might differ substantially from the temperature $T$ of the VS~\cite{Ackerman:2008kmp,Feng:2008mu}. In this work, we do not make any assumptions on how this difference in temperature arises in the early Universe. This question has, however, been addressed in previous works, see \cite{Berezhiani:1995am,Adshead:2016xxj,Hardy:2017wkr,Chu:2011be}. One natural possibility is to assume distinct couplings of the HS and VS to the inflaton, a class of scenarios called \textit{asymmetric reheating}.

Here, we revisit and expand on various aspects of a minimal yet instructive scenario in which the HS consists of self-interacting real scalar particles, which start out in thermal equilibrium with temperature $\Td$, but later undergo chemical decoupling, once their self-interaction rate drops below the Hubble rate. 
This model, with Lagrangian 
\begin{align}
\mathcal{L}_\text{HS} = \frac12 \partial_\mu S\partial^\mu S - \frac 12 \mu^2 S^2 - \frac{\lambda}{4!}S^4\eqsp,
\label{eq:lag}
\end{align}
has already been considered in the literature, either with a fundamental scalar or from the standpoint of an effective theory. For instance, the above setup has been studied in~\cite{Bernal:2015xba,Arcadi:2019oxh} for $\mu^2>0$, i.e.~for the case in which the $\mathbb{Z}_2$ symmetry is manifest.
In contrast, other studies~\cite{Ghosh:2022asg} expand on these results by considering an explicit breaking of the $\mathbb{Z}_2$ symmetry via an additional cubic term $\mathcal{L} \supset \kappa S^3$. In both of these cases, the DM relic abundance is driven by $4\rightarrow2$ and/or $3\rightarrow2$ annihilation processes, instead of the usual $2\rightarrow2$ processes that are relevant for standard WIMP DM. There already exists a vast literature on such setups, in which the DM particles partially destroy each other (see e.g.~\cite{Carlson:1992fn, Hochberg:2014dra,Bernal:2015ova,Agashe:2014yua,Farina:2016llk,Soni:2016gzf}). Such candidates have been called \textit{cannibal DM}~\cite{Carlson:1992fn} or \textit{SIMP DM}~\cite{Hochberg:2014dra}. We adopt the former appellation in this work.

Now, in contrast to~\cite{Bernal:2015xba,Arcadi:2019oxh,Ghosh:2022asg}, we mainly, but not exclusively, focus on the case in which the $\mathbb{Z}_2$ symmetry is \textit{spontaneously} broken, i.e.~$\mu^2 < 0$, in which case the HS freeze-out can happen in the spontaneous symmetry breaking (SSB) phase. Since this might lead to rapid decays of the DM particles into SM states, we need to additionally assume that the Higgs portal~\cite{Patt:2006fw} coupling, say $\epsilon$ with $\mathcal{L} \supset  - \frac{\epsilon}{2}S^2 \vert H \vert^2$, is sufficiently small to evade current constraints on decaying DM.\footnote{Global symmetries are thought to be accidental and, at the least, to be broken explicitly by quantum gravity effects~\cite{Kallosh:1995hi,Witten:2017hdv}. Here we consider the simple case of spontaneous symmetry breaking but, either way, our DM candidate is eventually unstable.} Thus, similar to the setup in~\cite{Arcadi:2019oxh,Ghosh:2022asg}, we assume that the HS and VS are essentially secluded. This is technically natural, as the two sectors are completely disconnected in the limit $\epsilon \rightarrow 0$~\cite{Luty:2005sn}. Such a scenario may be motivated in different ways. First, a decaying DM particle could lead to observable signatures which might be testable in the future. Also, an analysis of the SSB phase ties into the recent interest for phase transitions in a HS, in particular with regard to the production of gravitational waves (GW)~\cite{Schwaller:2015tja,Caprini:2018mtu}. Due to its $\mathbb{Z}_2$ symmetry, the present model is in the same universality class as the Ising model, which generically features a second-order phase transition (see e.g.~\cite{Berges:2000ew}). However, more complex HSs can have a strong first-order phase transition with GW signatures. Finally, SSB with a discrete symmetry may produce domain walls (DW), which should eventually be unstable,  leading to the production of GW as well as non-thermal production of DM~\cite{Vilenkin:1984ib,Saikawa:2017hiv} (see e.g.~\cite{Ramazanov:2021eya} for a scenario that features all the aforementioned effects). Such an instability may, for instance, be driven by an explicit but tiny $\mathbb{Z}_2$ symmetry breaking cubic self-coupling, as alluded to above, or through a mixing with the SM via the Higgs portal. In such scenarios, it is important to determine the amount of DM that is produced by thermal cannibal processes.

Considering the Lagrangian from eq.~\eqref{eq:lag} in the SSB phase, the scalar particle receives an additional trilinear coupling. Consequently, the relic abundance is mainly driven by $3 \rightarrow 2$ and $4 \rightarrow 2$ processes, both of which are parametrically related. This setup constitutes a non-trivial limit of scalar theories with generic trilinear coupling (cf.~in particular~\cite{Ghosh:2022asg}), since, in the case of SSB, the \textit{tree-level} amplitude for any process $k \rightarrow 2$ with $k > 2$ vanishes when the incoming particles are non-relativistic~\cite{Smith:1992rq,Erickcek:2021fsu}.
For the problem of DM, this implies that all reactions driving the freeze-out process are velocity-suppressed in the broken phase, and consequently extra care is required when inferring the relic abundance via the Boltzmann equations. More precisely, the usual approximation of a Maxwell-Boltzmann distribution is not necessarily justified anymore, which we take into account via a tailored formalism for calculating the DM relic abundance, which is easy to implement while still providing $\mathcal{O}(1\%)$ accuracy. Additionally, due to the vanishing tree-level cross-section, 1-loop corrections might become important. We further estimate this effect and highlight the affected part of parameter space. 

While we mainly focus on the case of SSB, for the sake of comparison and completeness, we also apply our formalism to the scenarios discussed in~\cite{Arcadi:2019oxh,Ghosh:2022asg}. Our work thus directly expands on existing results in the literature. In addition, we also report our results in the "domain of thermal dark matter candidates", analogous to the analysis performed in~\cite{Coy:2021ann} for WIMP-like DM. This domain, drawn in the plane of $\xi = \Td/T$ (at DM decoupling) and the DM mass ($m_\phi$ in this paper), is a general construction, which encompasses the available parameter space for DM candidates undergoing freeze-out from a thermal bath with temperature $\Td \neq T$. 
In particular, we derive novel analytical scaling relations that are satisfied by cannibal DM candidates within this plane. Specifically, we show that, for fixed coupling and non-relativistic freeze-out, contours of constant DM relic abundance within this domain follow the relation $\xi \propto m_{\phi}^{-\sfrac13 (1 + \kappa)}$ with $0 < \kappa \ll 1$.  Interestingly, the case $\xi \propto m_{\phi}^{-\sfrac13}$, i.e.~$\kappa = 0$, is characteristic for particles undergoing relativistic freeze-out~\cite{Hambye:2020lvy,Boehm:2004th}, meaning that cannibal DM particles undergoing non-relativistic freeze-out exhibit similar features than those particles, as they only feature a slightly steeper scaling. On this basis, we call $\kappa$ the "anomalous scaling" of cannibal DM. In fact, we find that the validity of this statement even goes beyond the models discussed in this work, as it generally holds true for a broad class of cannibal DM scenarios (albeit for different values of the anomalous scaling). Given this broader applicability, we thus effectively map out the subset of the full parameter space that can host cannibal DM scenarios not only in our scenario but also beyond.

Our work is structured as followed: In sec.~\ref{sec:setup} we set up the HS model. In sec.~\ref{sec:evolution} we study the evolution of the DM abundance in the HS via a specific set of Boltzmann equations. In sec.~\ref{sec:results}, we then present our results by \textit{(i)} showcasing those combinations of parameters that lead to the correct relic abundance, while also making explicit the scaling relation between the temperature ratio of the two sectors and the DM mass (sec.~\ref{sec:scaling}), \textit{(ii)} reporting our results in the domain of thermal DM candidates, while also calculating the relevant cosmological and astrophysical constraints (sec.~\ref{sec:domain}), and \textit{(iii)} commenting on the potential implications of domain wall formation in our scenario (sec.~\ref{sec:domain_walls}). Finally, we conclude in sec.~\ref{sec:conclusions}, while some technical details of our calculation are delegated to the appendix.

\section{Hidden sector setup}
\label{sec:setup}

We will in particular focus on the specifics of the Lagrangian in~\eqref{eq:lag} with $\mu^2 < 0$, in which case the $\mathbb{Z}_2$ symmetry gets spontaneously broken due to a vacuum expectation value
\begin{align}
v = \sqrt{\frac{6\vert\mu^2\vert}{\lambda}}\eqsp.
\end{align}
After substituting $S = v + \phi$ into eq.~\eqref{eq:lag} in order to express the Lagrangian in terms of the DM field $\phi$, the Lagrangian takes the form
\begin{align}
\mathcal{L}_\text{HS} = \frac 12 (\partial_\mu \phi)(\partial^\mu \phi) - \frac 12 m_\phi^2 \phi^2 -  {\frac{\lambda v}{3!}} \phi^3 - \frac{\lambda}{4!} \phi^4
\label{eq:lag_sb}
\end{align}
with $m_\phi^2 \equiv 2 |\mu^2| = \lambda v^2/3$ being the mass squared of $\phi$.

The important peculiarity of this phase is the fact that the tree-level amplitude for any process $k\phi \rightarrow \phi\phi$ with $k > 2$ vanishes at the kinematical threshold, i.e.~if all initial-state particles have vanishing momenta. In fact, the vanishing of threshold amplitudes at tree-level is not specific to the Lagrangian in the SSB phase, but also occurs in the symmetric phase, albeit only for $k > 4$~\cite{Voloshin:1992nh}. 
To make this behaviour more explicit, let us consider a generalization of the Lagrangian in eq.~\eqref{eq:lag_sb} with some arbitrary cubic coupling $g$, which is not necessarily generated via symmetry breaking (cf.~\cite{Ghosh:2022asg}),
\begin{align}
\mathcal{L}_\text{HS} = \frac 12 (\partial_\mu \phi)(\partial^\mu \phi) - \frac 12 m_\phi^2 \phi^2 -  {\frac{g m_\phi}{3!}} \phi^3 - \frac{\lambda}{4!} \phi^4\eqsp.
\label{eq:lag_gen}
\end{align}
In this case, the tree-level amplitude for the process $3\phi \rightarrow \phi\phi$ at threshold is given by\footnote{There has been some discrepancy in the literature regarding this result, see e.g.~\cite{Farina:2016llk,Ertas:2021xeh,Erickcek:2020wzd,Erickcek:2021fsu,Erickcek:2021fsu,Ghosh:2022asg} and references therein. If one assumes spontaneous instead of explicit symmetry breaking, some of these results do not feature a vanishing of the amplitude at threshold, which contradicts the general findings of~\cite{Brown:1992ay,Smith:1992rq} as already pointed out in~\cite{Erickcek:2021fsu,Ghosh:2022asg}.}
\begin{align}
\left| {\cal M}_{3\phi\rightarrow \phi\phi} \right| \overset{\text{th.}}{=} \frac{5g}{8m_\phi}\left| 3\lambda - g^2 \right|
\end{align}
with the corresponding cross-section at threshold (here, we include all symmetry factors for identical particles in the initial and final state, similar to~\cite{Arcadi:2019oxh}, see also~\cite{Erickcek:2021fsu})
\begin{align}
\langle \sigma_{3\phi \rightarrow \phi\phi} v^2 \rangle \overset{\text{th.}}{=} \frac{25\sqrt{5}g^2(3\lambda - g^2)^2}{2! 3! \times 12288 \pi m_\phi^5}\eqsp.
\end{align}
Consequently, we find that $\left| {\cal M}_{3\phi\rightarrow \phi\phi} \right|$ and thus $\langle \sigma_{3\phi \rightarrow \phi\phi} v^2 \rangle$ vanish at threshold if $g^2 = 3\lambda$, which coincides with the case in which the cubic coupling is generated via SSB. Similar statements are also true for processes with more particles in the initial state, meaning that the $s$-wave contribution of any tree-level process in the SSB phase that changes the number density of $\phi$ is loop-suppressed at threshold. This statement, has indeed been proven using only generic properties of tree-level amplitudes~\cite{Brown:1992ay}\footnote{While this result is certainly intriguing, there does not seem to be any underlying physical reason. Instead, it rather appears to be a mathematical coincidence (see e.g.~page 6 of~\cite{Smith:1992rq} for a short description of why the amplitude vanishes).} and it also applies to theories with continuous symmetries and thus with Goldstone modes~\cite{Brown:1992ay}. Such theories may have a strong first order phase transition (see e.g.~\cite{Pisarski:1983ms}), a feature of interest for possible GW signatures. 

Based on these results, it becomes apparent that we have to go beyond the simple $s$-wave amplitudes in order to determine the DM abundance in the SSB phase. Instead, we have to incorporate higher orders in the velocity by explicitly keeping the full momentum dependence of the annihilation cross-section. Thus, a full tree-level calculation of the annihilation cross-section is warranted. We do so by \textit{(i)} implementing the respective model in \textsc{FeynRules}~\cite{Alloul:2013bka}, and \textit{(ii)} using the generated UFO model~\cite{Degrande:2011ua} in \textsc{MadGraph5\_aMC@NLO}~\cite{Alwall:2014hca} to evaluate the cross-section, including the $k$-particle final-state phase-space.\footnote{More precisely, we use \textsc{MadGraph5} only to sample a reference cross-section $\bar{\sigma}_{\phi\phi\rightarrow k\phi}(E)$ for $\lambda=1$ and $\bar{m}_\phi=100\,\mathrm{MeV}$. Afterwards we simply rescale $\bar{\sigma}_{\phi\phi\rightarrow k\phi}(E)$ in order to obtain the full cross-section $\sigma_{\phi\phi\rightarrow k\phi}(E)$ for arbitrary masses and couplings, i.e.
\begin{align}
\sigma_{\phi\phi\rightarrow k\phi}(E) = \lambda^k \bar{\sigma}_{\phi\phi\rightarrow k\phi}(E \bar{m}_\phi/m_\phi) \times (\bar{m}_\phi/m_\phi)^2\nonumber\eqsp.
\end{align}
We explicitly checked that such a rescaling leads to errors below the percent level.}

However, since the tree-level amplitude vanishes at threshold, we also have to take into account that there might exist a non-negligible contribution from the 1-loop amplitude, whose value at threshold for the process $3\phi \rightarrow \phi\phi$ is of the order of~\footnote{We thank Camilo Garcia Cely for help regarding the estimation of the 1-loop amplitude.}
\begin{equation}
|\mathcal{M}_{3\phi\rightarrow \phi\phi}^\text{1-loop}| \overset{\text{th.}}{\simeq} \frac{81f \lambda^2}{2 \pi v}
\label{eq:M1l}
\end{equation}
with $f\sim \mathcal{O}(1)$. However, we find this contribution to be negligible as long as $\lambda \leq 10^{-1}$, while $\mathcal{O}(1\%)$ corrections arise for $\lambda = 1$. Hence, since the evaluation of the full momentum-dependent cross-section at 1-loop level is rather involved, we drop this contribution for all practical calculations, but later indicate the \textit{small} part of parameter space, which is potentially affected by this correction (cf.~fig.~\ref{fig:domain} in sec.~\ref{sec:domain}).

Additionally, we also have to make sure that the freeze-out in our model indeed happens in the SSB phase, i.e.~at temperatures when the $\mathbb{Z}_2$ symmetry is already broken. Since the DM particles in our setup usually decouple at $m_\phi/\Td \sim \mathcal{O}(1)$ (cf.~sec.~\ref{sec:evolution}), we thus have to require that the mass of the particle is well below the critical temperature $\Td_c$ at which the $\mathbb{Z}_2$ symmetry is restored, i.e.~$m_\phi \sim \Td \ll \Td_c$. A quick way to estimate $\Td_c$ given eq.~\eqref{eq:lag} is to use the thermal average $\langle S^2 \rangle_{\Td} \approx \Td^2/12$ at high $\Td$, which allows to approximate the finite $\Td$ effective potential as~\cite{Linde:1990flp}
\begin{align}
V_\text{HS}(\Td) \simeq \frac{\lambda}{2} \left (\frac{\Td^2}{24} - \frac{{\Td}^2_{\!\!\!c}}{24}\right) S^2  + \frac{\lambda}{4!} S^4
\label{eq:VHS}
\end{align}
with ${\Td}^2_{\!\!\!c} = 24 |\mu^2|/\lambda$. Thus, in order to assure that the freeze-out happens in the SSB phase, we have to enforce $m_\phi \ll \Td_c$, and consequently $\lambda \ll {\cal O}(10)$. This parameter range coincides well with the region in which 1-loop corrections become relevant, i.e.~$\lambda > 1$ (see above), meaning that we do not have to impose any additional restriction onto our analysis. Nevertheless, we will later indicate the region of parameter space, in which SSB happens before freeze-out (cf.~fig.~\ref{fig:domain} in sec.~\ref{sec:domain}).

For future reference, let us finally note that we also compare our results to those without SSB (cf.~\cite{Arcadi:2019oxh}), for which we employ the Lagrangian
\begin{align}
\mathcal{L}_\text{HS} = \frac 12 (\partial_\mu \phi)(\partial^\mu \phi) - \frac 12 m_\phi^2 \phi^2 - \frac{\lambda}{4!} \phi^4
\label{eq:lag_nsb}
\end{align}
instead of eq.~\eqref{eq:lag} in order to unify the notation.

\section{Evolution of the hidden sector}\label{sec:evolution}

Given the HS setup from the previous section, we now quantify the thermal evolution of the so-defined HS, in order to determine the combination of parameters that yields the correct relic abundance. While we mainly focus on the model defined via eq.~\eqref{eq:lag_sb}, the formalism discussed below remains equally valid for other scenarios, e.g.~those defined via eqs.~\eqref{eq:lag_gen} and~\eqref{eq:lag_nsb}, given that the correct interactions are taken into account.

For sufficiently large annihilation cross-sections\footnote{We will later indicate the region of parameter space, where the assumption of initial chemical equilibrium is no longer justified (cf.~fig.~\ref{fig:domain} in sec.~\ref{sec:domain}).} and at high temperatures, the interactions induced by eq.~\eqref{eq:lag_sb} ensure that the DM particle is in chemical and kinetic equilibrium with itself via reactions of the form $\phi \phi \leftrightarrow k  \phi$ with $k \geq 2$. However, once the temperature of this system approaches the mass of the scalar, number-changing interactions with $k > 2$ start to become inefficient, which leads to the chemical decoupling of $\phi$. Nevertheless, throughout this process, the reaction $\phi \phi \leftrightarrow \phi \phi$ remains efficient enough to keep $\phi$ in kinetic equilibrium, meaning that the evolution of $\phi$ can still be fully described by a Bose-Einstein (BE) distribution with HS temperature $\Td$ and chemical potential $\mu_\phi$,\footnote{We assume the Bose-Einstein distribution to be valid also for interacting scalars (see e.g. page 86 in ~\cite{alma991004090129703414}), including for large, but still pertubative, couplings.}
\begin{align}
    f_\phi(E_\phi, t) = \frac{1}{e^{[E_\phi - \mu_\phi(t)]/\Td(t)}-1}\eqsp.
    \label{eq:spectrum}
\end{align}
In order to determine the evolution of the phase-space distribution $f_\phi$, we thus have to formulate a set of two differential equations for the dynamical quantities $\Td(t)$ and $\mu_\phi(t)$. To this end, we can employ \textit{(i)} energy conservation in the HS, $\nabla_\mu T^{0\mu}_\phi = 0$, which leads to an equation of the form
\begin{align}
\dot{\rho}_\phi &+ 3H(\rho_\phi + P_\phi) = 0\label{eq:rhophi}
\end{align}
for the energy density $\rho_\phi$ and the pressure $P_\phi$ of $\phi$ with the Hubble rate $H$, as well as \textit{(ii)} the Boltzmann equation for the number density $n_\phi$ of $\phi$ (also cf.~\cite{Arcadi:2019oxh}), i.e.
\begin{align}
\dot{n}_\phi &+ 3Hn_\phi = \sum_{k>2} (k-2) \mathcal{C}_{\phi\phi \rightarrow k\phi}\left( 1- e^{(k-2)\mu_\phi/\Td} \right)\eqsp.\label{eq:nphi}
\end{align}
In the latter expression, $\mathcal{C}_{\phi\phi \rightarrow k\phi}$ is the integrated collision operator for the reaction $\phi\phi \rightarrow k \phi$,
\begin{align}
\mathcal{C}_{\phi \phi \rightarrow k\phi} & = \int_{\mathbb{R}^{3(k+2)}} (2\pi)^4|\mathcal{M}_{\phi \phi \rightarrow k\phi}|^2 \delta\left(\underline{p}_1 + \underline{p}_2 - \sum_{l=1}^{k}\underline{p}_{l+2}\right) \nonumber \\
& \hspace{1cm} \times f_\phi(E_1) f_\phi(E_2) \prod_{i=1}^{k} \left[1+f_\phi(E_{i+2})\right] \prod_{j=1}^{k+2} \text{d}\Pi_j \label{eq:cfull}
\end{align}
with $\text{d}\Pi_j = \text{d}^3 p_j/[(2\pi)^3 2E_j]$. Moreover, the factor $(k-2)$ encodes the number of net particles produced, $\phi$ is assumed to have $g_\phi=1$ degrees of freedom, and we have further used the fact that detailed balance implies $\mathcal{C}_{k\phi \rightarrow \phi\phi} = -e^{(k-2)\mu_\phi/\Td} \mathcal{C}_{\phi\phi \rightarrow k\phi}$ for the collision operator $\mathcal{C}_{k\phi \rightarrow \phi\phi}$ of the reverse reaction in thermal equilibrium.\footnote{Note that this relation is true for arbitrary spin-statistics and does not assume a Maxwell-Boltzmann statistics for $\phi$.} Since $\phi$ stays in kinetic equilibrium throughout the freeze-out process, all cosmological quantities $Q\in\{ n_\phi, \rho_\phi, P_\phi \}$ can be interpreted as functions of only $\Td$ and $\mu_\phi$ via their dependence on the distribution function $f_\phi$ from eq.~\eqref{eq:spectrum}. Consequently, by using the relation
\begin{align}
\dot{Q} = \frac{\partial Q}{\partial \Td}\dot{\Td} +  \frac{\partial Q}{\partial \mu_\phi}\dot{\mu}_\phi\eqsp,
\end{align}
eqs.~\eqref{eq:rhophi} and~\eqref{eq:nphi} can be transformed into a set of differential equations for $\Td(t)$ and $\mu_\phi(t)$~\cite{Bringmann:2020mgx}, which can be solved numerically for a given Hubble rate $H$, as well as a set of matrix elements $|\mathcal{M}_{\phi\phi\rightarrow k\phi}|$ that are used to calculate the collision operators $\mathcal{C}_{\phi\phi \rightarrow k\phi}$.

Notably, however, while we consider the HS in this setup to be fully decoupled from the VS, the Hubble rate $H$ and thus eqs.~\eqref{eq:rhophi} and \eqref{eq:nphi} still indirectly depend on the particle content of the VS as $H^2 = 8\pi G/3 (\rho_\phi + \rho_\text{VS})$.
Here, $\rho_\text{VS}$ is the energy density of the degrees of freedom in the VS, e.g.~the SM, which generally depends on both the photon temperature $T$ and the neutrino temperature $T_\nu$, with $T_\nu \neq T$ after neutrino decoupling. Consequently, in order to close the system, eqs.~\eqref{eq:rhophi} and \eqref{eq:nphi} need to be complemented by two additional equations describing the dynamics of $T(t)$ and $T_\nu(t)$. To this end, we assume instantaneous neutrino decoupling and use energy-momentum conservation in the VS to obtain~\cite{Hufnagel:2018bjp,Depta:2020zbh}
\begin{align}
\dot{T} = -\frac{3H(\rho_\text{VS} + P_\text{VS})}{\text{d} \rho_\text{VS}/\text{d}T}\eqsp, \quad \dot{T}_\nu = \dot{T}\big|_{T=T_\nu}
\label{eq:before_nud}
\end{align}
\textit{before} neutrino decoupling, and
\begin{align}
\dot{T} = -\frac{3H(\rho_{\text{VS},\slashed{\nu}} + P_{\text{VS},\slashed{\nu}})}{\text{d} \rho_{\text{VS},\slashed{\nu}}/\text{d}T}\eqsp,  \quad \dot{T}_\nu = -HT_\nu
\label{eq:after_nud}
\end{align}
\textit{after} neutrino decoupling. Here, $P_\text{VS}$ is the total pressure of the VS, while $\rho_{\text{VS},\slashed{\nu}}$ and $P_{\text{VS},\slashed{\nu}}$ denote the energy density and the pressure of the VS without neutrinos. Eqs.~\eqref{eq:before_nud} and~\eqref{eq:after_nud} are separated by the neutrino-decoupling temperature $T_{\nu\text{d}}$, which we calculate by following~\cite{Hufnagel:2018bjp,Depta:2020zbh}, i.e.~by solving $T_{\nu\text{d}}^5/T_{\nu\text{d},\text{SM}}^5 = H(T_{\nu\text{d}})/H(T_{\nu\text{d}, \text{SM}})$ with the neutrino-decoupling temperature $T_{\nu\text{d},\text{SM}} \approx 1.4\,\mathrm{MeV}$~\cite{Dolgov:2002wy,Bennett:2019ewm} in the SM. Overall, our considerations thus lead to a set of four coupled differential equations for the quantities $\Td(t)$, $\mu_\phi(t)$, $T(t)$, and $T_\nu(t)$.

Regarding the boundary conditions for these equations, the initial chemical potential is fixed to $\mu_\phi = 0$ at early times, since detailed balance implies $(k-2)\mu_\phi = 0$ and hence $\mu_\phi = 0$ as long as any reaction $\phi\phi \leftrightarrow k \phi$ with $k > 2$ is in equilibrium. The initial temperature, however, is a free parameter, which we express in terms of the ratio $\xi_\infty$ between the HS temperature $\Td$ and the SM/VS temperature $T$ in the limit $T\rightarrow \infty$,
\begin{align}
    \xi_\infty \equiv \lim_{T\rightarrow \infty}\left(\frac{\Td(T)}{T}\right)\eqsp,
\end{align}
i.e.~at early times when all particles are still relativistic. Note that typically $\xi_\infty \ll 1$  for the kind of DM that we consider in this work.

Finally, let us note that eqs.~\eqref{eq:rhophi} and \eqref{eq:nphi} do not make any other assumptions besides the one that $\phi$ obeys BE statistics, meaning that they fully describe the evolution of $\phi$ as long as kinetic equilibrium is maintained. However, the evaluation of the full collision operator in eq.~\eqref{eq:cfull} is rather involved. One simplification is to only use Maxwell-Boltzmann (MB) statistics\footnote{MB statistics is only used to approximate $\mathcal{C}_{\phi \phi \rightarrow k\phi}$, and we still use the full BE distribution in the calculation of $n_\phi$ etc.} for $f_\phi$ in $\mathcal{C}_{\phi \phi \rightarrow k\phi}$, which yields~\cite{Arcadi:2019oxh}
\begin{align}
    \mathcal{C}_{\phi \phi \rightarrow k\phi} &\simeq \frac{2\Td e^{2\mu_\phi/\Td}}{\pi^4} \int_{m_\phi}^\infty \sigma_{\phi\phi\rightarrow k\phi}(E)\times E^2(E^2-m_\phi^2)K_1(2E/\Td)\,\text{d}E\label{eq:coll}
\end{align}
with the cross-section $\sigma_{\phi\phi\rightarrow k\phi}$ for the process $\phi\phi\rightarrow k\phi$. This approach is justified, since the exact form of the collision operator becomes relevant only close to chemical decoupling, which happens at $m_\phi/\Td \sim \mathcal{O}(1)$, i.e.~when $\phi$ is already semi- or non-relativistic. We will later compare our results for the case without SSB to those of~\cite{Arcadi:2019oxh} (cf.~figure~\ref{fig:comp}), which have been calculated with full BE statistics for $f_\phi$ in $\mathcal{C}_{\phi \phi \rightarrow k\phi}$, and show that our approximation usually leads to errors at the $\mathcal{O}(1\%)$ level, which we deem perfectly sufficient for our purposes. Hence, in the following, we use eq.~\eqref{eq:coll} as a proxy for $\mathcal{C}_{\phi \phi \rightarrow k\phi}$ in eq.~\eqref{eq:nphi}, while keeping the full BE statistics everywhere else. Finally, let us note that, for the Lagrangian in eq.~\eqref{eq:lag_sb}, we only consider $k\in\{3, 4\}$ in eq.~\eqref{eq:nphi}, since processes with more final-state particles are severely suppressed and hence negligible (see below). In fact, even the inclusion of $k=4$ merely leads to small corrections compared to only using $k=3$.

\begin{figure*}[t]
    \centering
    \begin{center}
    \includegraphics[width=0.495\textwidth]{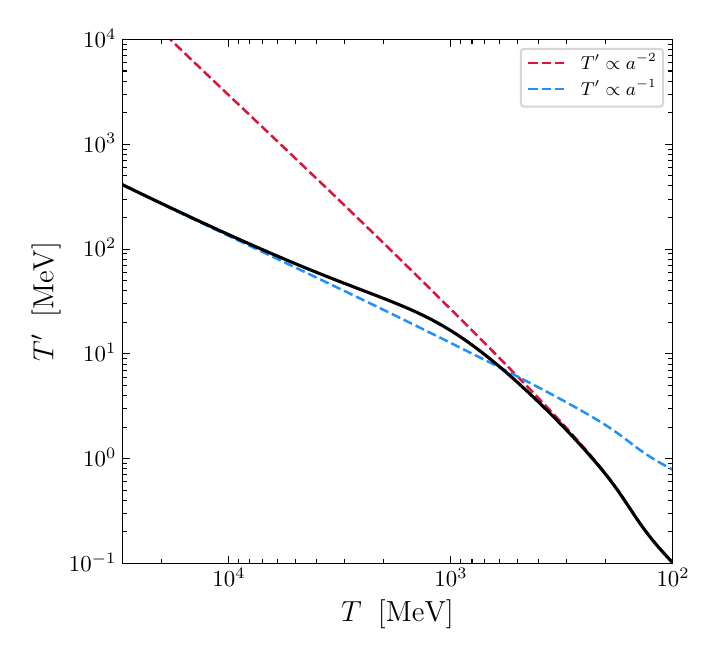}
    \includegraphics[width=0.495\textwidth]{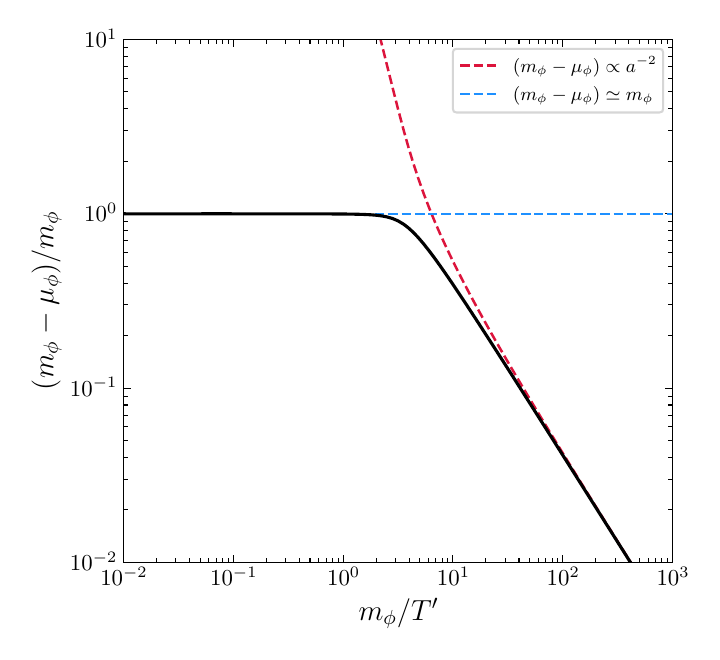}
    \end{center}
    \caption{\textit{Left:} Evolution of the dark-sector temperature $\Td$ as a function of the SM temperature $T$ (black) for the benchmark point discussed in the text,  $m_\phi = 100\,\mathrm{MeV}$, $\lambda = 10^{-2}$, and $\xi_\infty = 1.44\times 10^{-2}$. For reference, we also indicated the expected behaviour in the ultra-relativistic ($\Td \propto a^{-1}$, dashed blue) and non-relativistic ($\Td \propto a^{-2}$, dashed red) regime. \textit{Right:} Evolution of the chemical potential $\mu_\phi$ as a function of $m_\phi/\Td$ for the same benchmark point. Again, the expected behaviour in the ultra-relativistic ($m_\phi - \mu_\phi \sim m_\phi$, dashed blue) and non-relativistic ($m_\phi - \mu_\phi \propto a^{-2}$, dashed red) regime is also indicated.}
    \label{fig:Td_muphi}
\end{figure*}

To illustrate the general solution of the above equations, in fig.~\ref{fig:Td_muphi} we show the exemplary evolution of $\Td$ and $\mu_\phi$ for the Lagrangian in eq.~\eqref{eq:lag_sb} with $m_\phi = 100\,\mathrm{MeV}$, $\lambda = 10^{-2}$, and $\xi_\infty = 1.44\times 10^{-2}$. This combination is chosen in such a way that $\phi$ assumes the correct relic abundance to account for all of DM, i.e.~$\Omega_\phi h^2 = 0.12$~\cite{ParticleDataGroup:2020ssz}. In the left panel, the dark-sector temperature $\Td$ (solid black) is illustrated as a function of the SM temperature $T$. Before decoupling at $m_\phi/\Td \sim \mathcal{O}(1)$, $\phi$ is still highly relativistic and $\Td \propto a^{-1}$ (cf.~dashed blue line) with the scale factor $a$ fulfilling $H = \dot{a}/a$, in accordance with the usual scaling for a thermally-coupled, ultra-relativistic particle~\cite{Kolb:1990vq}. Once $\phi$ starts to become non-relativistic -- but before chemical decoupling -- $\phi$ enters a cannibal phase~\cite{Pappadopulo:2016pkp}. In this regime, entropy conservation in the dark-sector implies $m_\phi^3 e^{-m_\phi/\Td}\sqrt{\Td/m_\phi}a^3 = \text{const}$~\cite{Pappadopulo:2016pkp} and hence $\Td \sim \log(a)^{-1}$, i.e.~the dark-sector temperature enters a period of slower cooling, as the processes $k\phi\rightarrow \phi \phi$ with $k > 2$ start to efficiently transform rest mass into kinetic energy. After chemical decoupling, however, these reactions become inefficient and cannibalism comes to a natural close. Afterwards, $\Td$ is subject to only redshift, and we recover the usual relation $\Td \propto a^{-2}$ for a decoupled, non-relativistic particle (cf.~dashed red line)~\cite{Kolb:1990vq}. Additionally, in the right panel, we show the corresponding chemical potential $\mu_\phi$ (solid black) as a function of $m_\phi/\Td$. At early times, the chemical potential vanishes, i.e.~$(m_\phi-\mu_\phi) \sim m_\phi$ (cf. dashed blue line), since number-changing processes are still in equilibrium. However, once these reactions become inefficient, i.e.~during chemical decoupling at $m_\phi/\Td \sim \mathcal{O}(1)$, $\mu_\phi$ starts to increase in order to counteract the Boltzmann-suppression. After freeze-out has concluded, this finally leads to the asymptotic scaling $(m_\phi-\mu_\phi) \propto a^{-2}$ (cf.~dashed red line) and hence $(m_\phi-\mu_\phi)/\Td = \text{const}$, which also follows from the conservation of the comoving number density, $n_\phi \propto a^{-3}$, since $n_\phi \propto \Td^{3/2}e^{-(m_\phi-\mu_\phi)/\Td}$ and $\Td \propto a^{-2}$ (see above).

\begin{figure*}[t]
    \centering
    \begin{center}
    \includegraphics[width=0.495\textwidth]{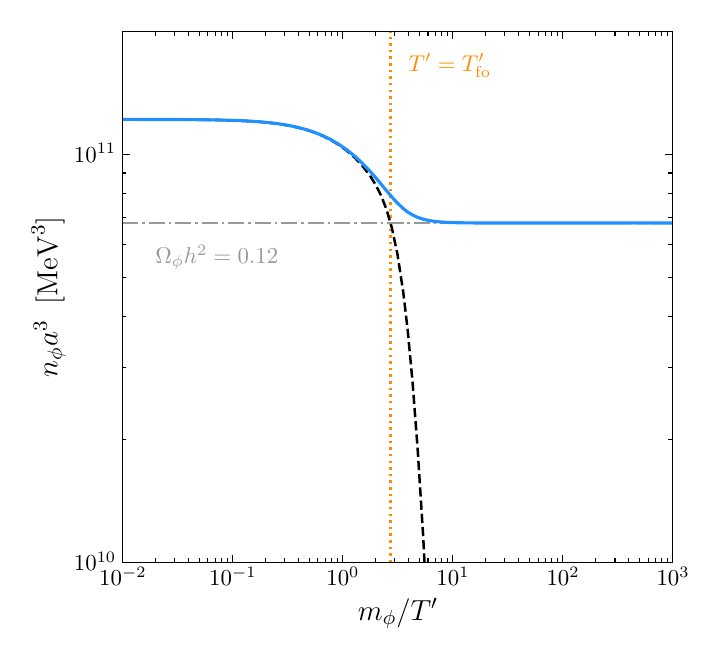}
    \includegraphics[width=0.495\textwidth]{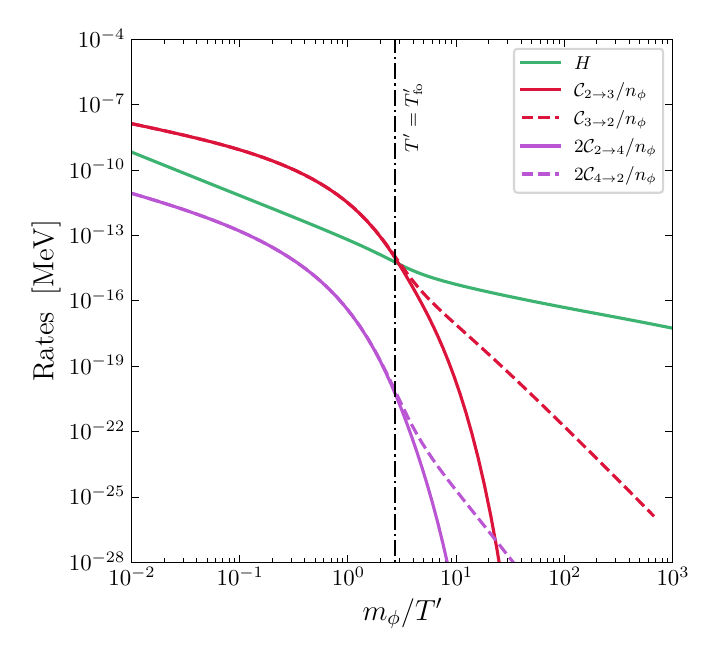}
    \end{center}
    \caption{\textit{Left:} Evolution of the abundance $n_\phi a^3$ (solid green) as a function of $m_\phi/\Td$ for the same benchmark point as in fig.\ref{fig:Td_muphi} and specified in the text. Additionally, the thermal value $\bar{n}_\phi a^3$ (dashed black) as well as the value corresponding to the correct relic abundance (dash-dotted grey) are also indicated, as well as a proxy for the decoupling temperature $\Td_\text{fo}$ (dotted orange) \textit{Right:} Comparison of the Hubble rate (solid green) with the forward (dashed) and backward (solid) reactions rates for the processes $\phi\phi\rightarrow 3\phi$ (red) and $\phi\phi\rightarrow 4\phi$ (purple), all as a function of $m_\phi/\Td$.}
    \label{fig:nphi}
\end{figure*}

For comparison, in fig.~\ref{fig:nphi} we further show the comoving number density $n_\phi a^3$ (left), as well as a comparison between the Hubble rate and the relevant interaction rates (right) as a function of $m_\phi/\Td$. Overall, $n_\phi a^3$ (solid blue, left) shows the expected behaviour of a particle freezing-out from equilibrium: At large temperatures, $\Td \gg m_\phi$, $n_\phi a^3$ follows its value in chemical equilibrium (dashed black, left), corresponding to $\mu_\phi = 0$, but later starts to deviate from it at around $\Td \sim m_\phi/\mathcal{O}(1)$ as the reactions maintaining equilibrium fall below the Hubble rate (cf. right panel of fig.~\ref{fig:nphi}). This way, $\phi$ maintains a non-vanishing relic abundance, which, by choice of parameters, matches the observed DM abundance today (dashed-dotted grey, left). To quantify the decoupling process, we can define the temperature at freeze-out $\Td_\text{fo}$ as the intersection between the black and grey line, corresponding to instantaneous decoupling. For the current benchmark point, we find $x_\text{fo}' \equiv m_\phi/\Td_\text{fo} \sim 3$, which differs substantially from the usual WIMP case with $m_\text{DM}/T_\text{fo} \sim 20$. This is because the collision operators $\mathcal{C}_{\phi \phi \rightarrow k\phi} \simeq \langle \sigma_{\phi \phi \rightarrow k\phi} v^{k-1}\rangle n_\phi^k$ do not only get suppressed once $\phi$ becomes non-relativistic at $\Td \sim m_\phi$, but already earlier, as the cross-section $\sigma_{\phi \phi \rightarrow k\phi}$ vanishes at $\Td \sim \sqrt{s} = (k/2)m_\phi$, when the two initial-state particles no longer have enough kinetic energy to produce $k$ final-state particles.\footnote{Since $\mathcal{C}_{4\phi \rightarrow \phi\phi}$ ($\mathcal{C}_{3\phi \rightarrow \phi\phi}$) gets suppressed at $\Td~\sim~2m_\phi$ ($\Td~\sim~1.5m_\phi$), the reaction $3\phi \rightarrow \phi\phi$ always stays in equilibrium a little longer, meaning that the reaction $4\phi \rightarrow \phi\phi$ can essentially be neglected. This is especially true for $\lambda \ll 1$.}
Since $(k/2)m_\phi > m_\phi$ for $k \geq 3$, this implies comparatively large couplings are required for non-relativistic freeze-out.

\section{Results}
\label{sec:results}

\begin{figure*}[t]
    \centering
    \begin{center}
    \includegraphics[width=0.495\textwidth]{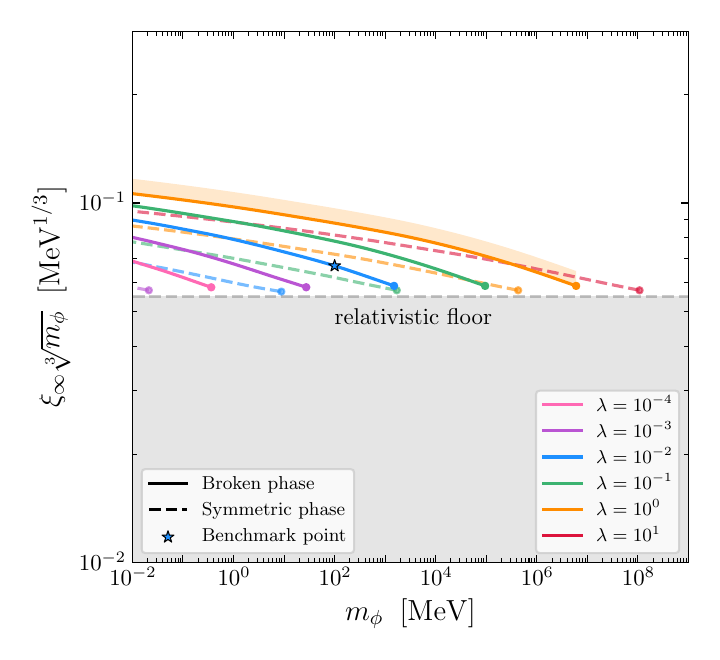}
    \includegraphics[width=0.495\textwidth]{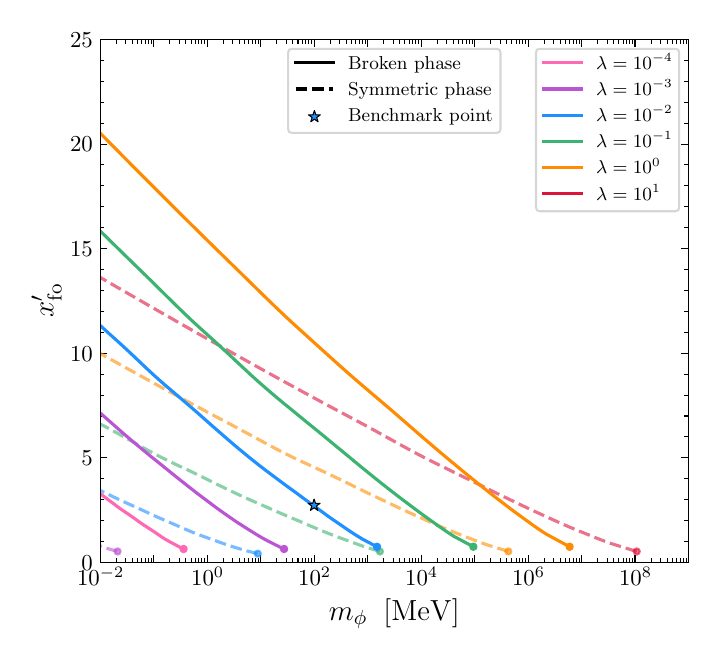}
    \end{center}
    \caption{\textit{Left:} Contours of constant $\lambda$ (different colours) to produce the correct relic abundance in the $\xi_\infty\sqrt[3]{m_\phi}-m_\phi$ parameter plane with (solid) and without (dashed) SSB. We further mark the points beyond which the HS is not initially in chemical equilibrium (filled circles), the effects of potential one-loop corrections in the case with SSB (shaded band for $\lambda=1$), as well as the benchmark point that was used for fig.~\ref{fig:Td_muphi} and \ref{fig:nphi} (blue star). \textit{Right:} The parameter $x_\text{fo}' = m_\phi/\Td_\text{fo}$ as a function of $m_\phi$ for the same couplings and scenarios as in the left panel.}
    \label{fig:contours}
\end{figure*}

Given the evolution of the HS, we now fix $\lambda$ via the condition $\Omega_\phi h^2 = 0.12$~\cite{Aghanim:2018eyx} and afterwards plot the resulting contours of constant coupling in the $\xi_\infty \sqrt[3]{m_\phi} - m_\phi$ parameter plane. This way, it is possible to map out the locus of all DM candidates that match the observed dark-matter abundance in our scenario. The results of this procedure for the case with SSB according to the Lagrangian in eq.~\eqref{eq:lag_sb} (solid lines) are shown in the left panel of fig.~\ref{fig:contours}, while -- for comparison -- we also provide the results for the case without SSB according to the Lagrangian in eq.~\eqref{eq:lag_nsb} (dashed lines). In both cases, the different colours encode different values of the quartic coupling $\lambda$, ranging from $\lambda = 10^{-4}$ (pink) to $\lambda=10$ (red).\footnote{In addition to the considerations at the end of sec.~\ref{sec:setup}, we take the usually quoted value $\lambda \lesssim 4 \pi \sim 10$ as maximal perturbative coupling.} Note that, when considering SSB, we do not show the line for $\lambda=10$, as in this case the freeze-out would happen before symmetry breaking (cf.~the discussion around eq.~\eqref{eq:VHS}). For smaller couplings, $\lambda \sim 1$, 1-loop corrections might become important according to eq.~\eqref{eq:M1l}, and we indicate the effect of such potential corrections as an uncertainty band around the respective line. Additionally, for each value of $\lambda$, there exists a maximal value of $m_\phi$ (filled circle), beyond which the HS is not initially in equilibrium. Consequently, larger masses spoil the validity of eq.~\eqref{eq:spectrum} and thus require a more detailed calculation. Finally, we also indicate the benchmark point (blue star) that was used in sec.~\ref{sec:evolution} for figs.~\ref{fig:Td_muphi} and~\ref{fig:nphi}.

In general, there exists a lower bound on $\xi_\infty$ (dashed grey), which is set by those DM particles that decouple from the thermal bath while being relativistic. In this case, $\Omega_\phi h^2 \propto m_\phi \xi_\infty^3$, meaning that $\Omega_\phi h^2 = \text{const}$ implies $\xi_\infty \propto m_\phi^{-\sfrac13}$. This corresponds to a horizontal line in the $\xi_\infty \sqrt[3]{m_\phi}- m_\phi$ parameter plane, which has previously been dubbed \textit{the relativistic floor}~\cite{Hambye:2020lvy,Coy:2021ann}. Consequently, for temperature ratios below this line, $\phi$ can never account for all the DM abundance in the Universe, thus making this region inaccessible.

\subsection{Scaling relations between $m_\phi$ and $\xi_\infty$}
\label{sec:scaling}

For temperature ratios above the floor, we find that the resulting contours of constant $\lambda$ can only be understood numerically if the freeze-out happens in the ultra- to semi-relativistic regime. On the contrary, however, it is possible to quantify their behaviour analytically in the case of non-relativistic decoupling, i.e.~far away from the relativistic floor. While these are not the first analytical results regarding the freeze-out of cannibal DM, they -- for the first time -- precisely quantify the relation between the DM mass and the temperature of the HS.

Assuming for simplicity that freeze-out happens instantaneously, contours of constant relic abundance are defined by the relation
\begin{align}
\text{d}\left( m_\phi Y_\phi |_{T=T_\text{fo}} \right) = \text{d}\left( m_\phi \frac{n_\phi}{s_\text{VS}}\Big|_{T=T_\text{fo}} \right) = 0
\label{eq:const_oh2}
\end{align}
with $Y_\phi = n_\phi/s_\text{VS}$ and the entropy density $s_\text{VS}$ of the VS. In the non-relativistic limit, the entropy density $s_\phi$ of $\phi$ can further be approximated as $s_\phi \simeq \left(x_\text{fo}' + \sfrac52\right) n_\phi$, while separate entropy conservation in both sectors dictates $s_\phi/s_\text{VS} \propto \xi_\infty^3$ up to factors of $g_{*s}$, which is the number of entropy degrees of freedom in the VS. Hence, eq.~\eqref{eq:const_oh2} becomes
\begin{align}
\text{d} \left[ m_\phi \left(x_\text{fo}' + \frac52\right)^{-1} \xi_\infty^3 \right] \simeq 0\eqsp.
\end{align}
For $x_\text{fo}' = \text{const}$, this relation would imply that $\Omega_\phi h^2 = \text{const}$ is realized along curves parallel to the relativistic floor with $\xi_\infty \propto m_\phi^{-\sfrac13}$ -- independently of the annihilation cross-section. However, since $\partial_{m_\phi} x_\text{fo}',\partial_{\xi_\infty} x_\text{fo}' \neq 0$, this naive scaling relation necessarily receives model-dependent corrections. These corrections can be calculated on a rather general note, even beyond the model considered in this work, by assuming that at freeze-out the thermally averaged cross-section can be parameterized as\footnote{For a given model, the value of $k$ is then set by the multiplicity of the dominant number changing interaction, while $\beta=0$ is used for $s$-wave dominated processes and $\beta\neq 0$ otherwise.}
\begin{align}
\langle \sigma_{k} v^{k-1} \rangle \sim m_\phi^{-3k+4} x_\text{fo}'^{-\beta}
\label{eq:svk}
\end{align}
with some parameter $\beta$. Then, after equating $\langle \sigma_{k} v^{k-1} \rangle n_\phi^{k-1} \sim H \sim T_\text{fo}^2/M_\text{pl}$ with the Planck mass $M_\text{pl}$ to determine $x_\text{fo}'(m_\phi, \xi_\infty)$, we find that for the class of models whose cross-section at freeze-out can be parameterized by eq.~\eqref{eq:svk}, curves of constant relic abundance are solutions of the differential equation (cf.~eq.~\eqref{eq:dxi_dm_exp} and more generally app.~\ref{app:details} for more details regarding this calculation)\footnote{This result is different from the statement in eq.~(67) of~\cite{Arcadi:2019oxh}. However, by comparing our results with those in~\cite{Arcadi:2019oxh} (cf.~fig.~\ref{fig:comp}), we find that their scaling actually matches ours, despite their claim stating otherwise.}
\begin{align}
\frac{\text{d}\xi_\infty}{\text{d} m_\phi} \simeq -\frac13 \frac{\xi_\infty}{m_\phi}\left[ 1 + \frac{5}{3k-5}\times \frac{1}{x_\text{fo}'} + \mathcal{O}\left(\frac{1}{x_\text{fo}'^2} \right) \right]\eqsp.
\label{eq:scaling_main}
\end{align}
Hence, in the non-relativistic freeze-out regime, the naive scaling relation receives a small correction of $\mathcal{O}(x_\text{fo}'^{-1})$, with an explicit dependence on $k$ and thus on the cannibal process that sets the relic abundance. For each point in parameter space, we then approximately obtain
\begin{align}
\xi_\infty \propto m_\phi^{-\sfrac13 (1+\kappa)} \quad \text{with} \quad \kappa \simeq 5  (3k-5)^{-1} x_\text{fo}'^{-1}
\label{eq:anomalous}
\end{align}
and $k=3$ ($k=4$) in the broken (symmetric) phase. Interestingly, far away from the relativistic floor, i.e.~for small masses and large initial temperature ratios, the simpler scaling relation $\xi_\infty \propto m_\phi^{-\sfrac13}$ is assumed. This is because, for large $x_\text{fo}'$ we find (cf.~eq.~\eqref{eq:xfo})
\begin{align}
\frac{3k-5}{3} \times x_\text{fo}' \simeq \text{const.} - \log\left( \frac{m_\phi}{M_\text{pl}} \right) + 2\log(\xi_\infty)
\label{eq:xcd_main}
\end{align}
and thus $x_\text{fo}' \rightarrow \infty$ or equivalently $\kappa \rightarrow 0$ for $m_\phi/M_\text{pl} \rightarrow 0$ and/or $\xi_\infty \rightarrow \infty$. We demonstrate this behaviour in the right panel of fig.~\ref{fig:contours}, which shows $x_\text{fo}'$ as a function of $m_\phi$ for the same values of $\lambda$ as in the left panel. Consequently, every contour of constant $\lambda$ converges, albeit logarithmically, to a line parallel to the relativistic floor.

Interestingly, the results in the left panel of fig.~\ref{fig:contours} are further suitable for deducing the dark matter relic abundance in the case of non-relativistic decoupling. This is because, in this limit,
\begin{align}
Y_\phi(T_\text{fo}) \simeq \left(x_\text{fo}' + \frac52\right)^{-1} \frac{s_\phi}{s_\text{VS}} = \frac{\xi_\infty^3}{g_{*s}^\infty \left(x_\text{fo}' + \frac52\right)}
\end{align}
with $g_{*s}^\infty = g_{*s}(T\rightarrow \infty)$. Hence, for a given value of $\xi_\infty$, it is possible to fully deduce $Y_\phi$ from the knowledge of $x_\text{fo}'$.

Additionally, it is worth noting that compared to the symmetric phase (dashed), in the broken phase (solid) larger values of $\lambda$ are required to obtain the correct relic abundance for fixed $m_\phi$ and $\xi_\infty$. This is because of the previously discussed fact that the amplitude for the dominant process $3\phi\rightarrow \phi\phi$ is loop-suppressed at threshold in the broken phase, while the dominant process $4\phi \rightarrow \phi\phi$ in the symmetric phase does not exhibit such behaviour. Consequently, this additional suppression needs to be compensated for by larger values of $\lambda$.

\begin{figure}[t]
\center
\includegraphics[width=0.6\textwidth]{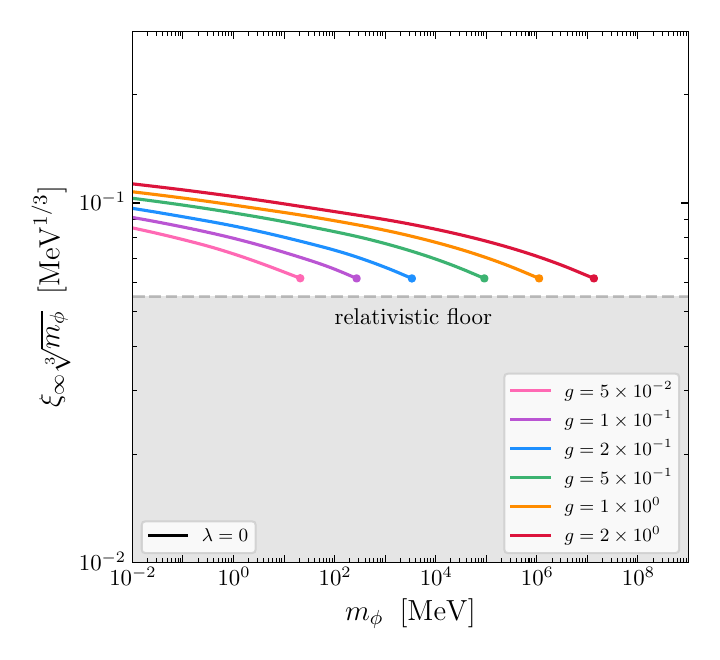}
\caption{Same as the left panel of fig.~\ref{fig:contours}, but for the model defined in eq.~\eqref{eq:lag_gen} with $\lambda=0$, i.e.~only a cubic coupling.}
\label{fig:gen}
\end{figure}

To further highlight the fact that the deduced behaviour is rather generic, in fig.~\ref{fig:gen} we also show the results of our procedure for the model defined in eq.~\eqref{eq:lag_gen} with $\lambda = 0$ and $g\neq0$ as an example, i.e.~a model with only a cubic coupling. Evidently, the behaviour remains similar, and the results fully respect the previously derived scaling relations, thus enforcing the generality of our results.

\subsection{Constraints in the $m_\phi-\xi_\mathrm{cd}$ plane}
\label{sec:domain}

\begin{figure*}[t]
    \centering
    \begin{center}
    \includegraphics[width=0.495\textwidth]{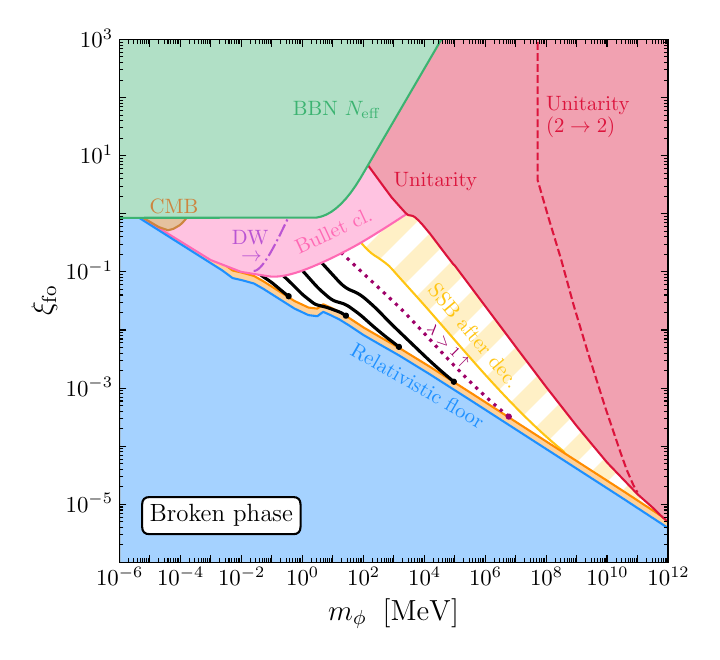}
    \includegraphics[width=0.495\textwidth]{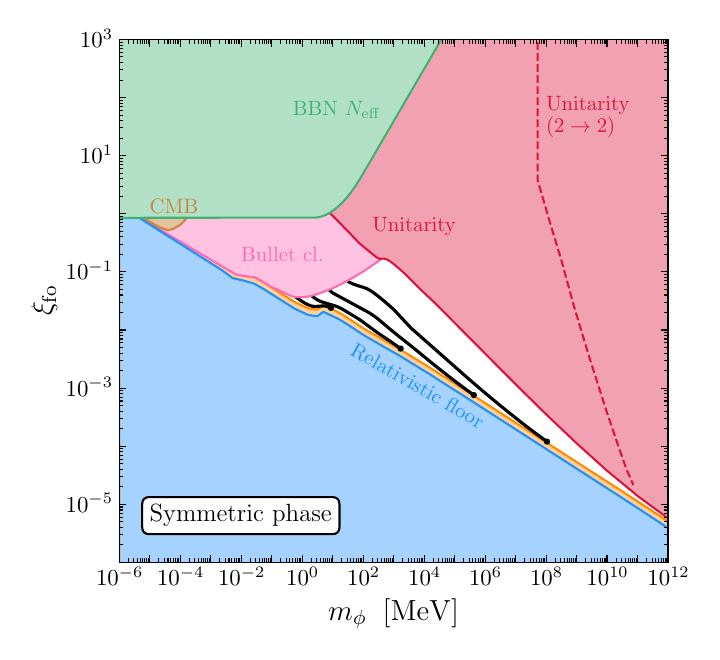}
    \end{center}
    \caption{\textit{Left:} Contours of constant relic abundance for the case with SSB and the same couplings as in fig.~\ref{fig:contours} (black lines). In the vein of~\cite{Coy:2021ann}, we also indicate the relativistic floor (blue, filled) as well as the other relevant constraints. In particular, we highlight those regions of parameter space that are excluded by a too large value of $N_\text{eff}$ at the time of BBN (green, filled), by measurements of the bullet cluster (pink, filled), and by measurements of the CMB power spectrum (brown, filled). Besides this, we also indicate those regions of parameter space that are generally inaccessible, either because HS thermalization is not feasible (orange, filled) or because the required cross-section would be beyond what is allowed by unitarity (red, filled). Finally, we also highlight the area, in which SSB happens after freeze-out (yellow, hatched) as well as the special case $\lambda = 1$ beyond which 1-loop corrections become important (purple, dotted line). \textit{Right:} Same as in the left panel, but for the symmetric phase.}
    \label{fig:domain}
\end{figure*}

We now report our results in the domain of thermal DM candidates, similar to what has been done in~\cite{Coy:2021ann}. To this end, we first translate our results into the $m_\phi-\xi_\text{fo}$ parameter plane, with $\xi_\text{fo}$ being the temperature ratio at decoupling. In general, the required translation $\xi_\infty \rightarrow \xi_\text{fo}$ is done numerically for each value of $m_\phi$. However, in the non-relativistic regime, we can also find an analytic approximation, since separate entropy conservation in both sectors implies
\begin{align}
\xi_\text{fo}^3 \simeq 6.9 \frac{g_{*s}(T_\text{fo})}{g_{*s}^\infty} \xi_\infty^3 x_\text{fo}^{-3/2}\left(x_\text{fo}' + \frac52\right)^{-1}e^{x_\text{fo}'}\eqsp.
\label{eq:xicd}
\end{align}
The resulting contours are shown in the left and right panel of fig.~\ref{fig:domain} for the case with and without SSB, respectively. Here the solid black lines correspond to the respective contours shown in the left panel of fig.~\ref{fig:contours}, with couplings getting larger by one order of magnitude each from left to right, starting with $\lambda = 10^{-4}$ in the broken and with $\lambda = 10^{-2}$ in the symmetric phase.\footnote{In the latter case, the lines for $\lambda \in \{10^{-4}, 10^{-3}\}$ are completely excluded and thus not visible.} In the broken phase, we also specifically indicate the line $\lambda = 1$ (purple, dotted line), above which \textit{1-loop corrections} become important (see the discussion above).

While the domain of possible cannibal DM candidates is bounded from below by the previously discussed \textit{relativistic floor} (blue, filled), there exist additional, complementary constraints that narrow down the viable parameter space even further.

First, the parameter space directly above the relativistic floor is inaccessible, since those combinations of parameters do not allow for \textit{HS thermalization} (orange, filled). This constraint exactly follows the solid black circles, which mark the point beyond which the HS is not initially in equilibrium (cf.~the discussion at the beginning of sec.~\ref{sec:results}).

Since the DM particle also features self-interactions of the form $\phi\phi\leftrightarrow \phi\phi$, there further exists a constraint on the transfer cross-section $\sigma_T$ from observations of the \textit{bullet cluster} (pink, filled). In this work, we employ the limit $\sigma_T/m_\phi < 1\,\text{cm}^2/\text{g}$~\cite{Randall:2008ppe} at a typical velocity of $v = 10^{-4}$. By calculating the full transfer cross-section
\begin{align}
\sigma_T = \int_{S_1} \frac{\text{d}\sigma_{\phi\phi\rightarrow\phi\phi}}{\text{d}\Omega} \times (1-\cos\theta)\;\text{d}\Omega 
\end{align}
in our scenario, we find
\begin{align}
m_\phi \lesssim 20.56\,\mathrm{MeV} \times \lambda^{2/3}\qquad(\text{broken phase})&\\
m_\phi \lesssim \textcolor{white}{2}8.16\,\mathrm{MeV} \times \lambda^{2/3} \!\!\!\quad(\text{symmetric phase})&\eqsp.
\end{align}
While these results indicate that the bound on the mass is weaker in the symmetric phase for $\lambda = 1$, for each point in the $m_\phi-\xi_\text{cd}$ parameter plane, the required value of $\lambda$ (to obtain the correct relic abundance) is larger in the case without symmetry breaking (cf.~left panel of fig.~\ref{fig:contours}). Overall, we thus find that the bullet-cluster constraint is stronger in the symmetric phase (cf.~fig.~\ref{fig:domain}).

In addition, some combinations of parameters are excluded as they would lead to cross-sections that go beyond the values allowed by \textit{unitarity} (red, filled). To quantify this effect, we solve the evolution equations again with the maximally allowed cross-section given in~\cite{Bhatia:2020itt}, i.e.
\begin{align}
\langle \sigma_{\phi\phi\rightarrow k\phi} v^{k-1} \rangle_\text{max} = \frac{2^{\frac{3k-2}{2}} (\pi x)^{\frac{3k-5}{2}}}{m_\phi^{3k-4}}
\end{align}
with $x = m_\phi/\Td$ and $k=3$ ($k=4$) in the broken (symmetric) phase. We then define the constraint as the line in $m_\phi-\xi_\text{cd}$ parameter space that leads to the correct relic abundance. For comparison, we also show the corresponding limit for $k=2$ in both cases (red dashed), which coincides with the limit calculated in~\cite{Coy:2021ann} for standard WIMPs. The constraint from unitarity is especially important, as it emphasizes the general assessment that viable cannibal DM candidates must reside closer to the relativistic floor than the usual WIMP candidates. This is because, the thermally-averaged annihilation cross-section gets suppressed already at $\Td \sim \sqrt{s} = (k/2)m_\phi$, meaning that comparatively large couplings are required to cause a large suppression in the number density (also cf.~sec.~\ref{sec:evolution}). In turn, these large couplings quickly go beyond the unitarity limit. Similar statements also hold true for cannibal DM models that are different from the one studied in this work, since the derivation of the scaling relation, specifically eq.~\eqref{eq:svk}, is generic. Thus, we can say on a general note that genuine, cannibal DM candidates cannot lie too far above the relativistic floor, as they would quickly get in conflict with constraints from unitarity.

In the broken phase, this statement is further fortified by the fact that \textit{SSB happens after freeze-out} (yellow, hatched) if $\lambda > 10$ (cf.~the discussion in sec.~\ref{sec:setup}), which pushes the available parameter space ever closer to the relativistic floor.

Other combinations of parameters are excluded since they would lead to a number of $N_\text{eff}$ that is too large at the time of \textit{big bang nucleosynthesis} (BBN) (green, filled). We calculate this constraint only roughly by demanding that the freeze-out of $\phi$ happens either before BBN, i.e.~$\Td_\text{cd} \lesssim 1\,\mathrm{MeV}$, or that its abundance at decoupling is still compatible with the current constraint $N_\text{eff} < 3.33$ at $2\sigma$~\cite{Aghanim:2018eyx}.\footnote{We find a minimally allowed value of $\xi_\text{cd}$ that is larger than the one shown in~\cite{Coy:2021ann}. This is because, the latter reference assumes three degrees of freedom, while we only have one, thus making our limits less stringent.} While this calculation could certainly be improved, a more thorough analysis would not change the overall results, since the BBN constraints are subdominant in all parts of parameter space.

Moreover, there exists a constraint from measurements of the \textit{cosmic microwave background (CMB)}~\cite{1992ApJ...398...43C,Buen-Abad:2018mas,Heimersheim:2020aoc} (brown, filled). Specifically, cannibal DM needs to be sufficiently non-relativistic at the time of photon decoupling, $T_\text{pd} \simeq 0.26\,\mathrm{eV}$, in order to behave like cold DM and thus not to suppress the matter power spectrum on small scales. This argument imposes a constraint on the HS temperature at the time of photon decoupling~\cite{Heimersheim:2020aoc}
\begin{align}
    \frac{\Td(T_\text{pd})}{m_\phi} < 10^{-5}\eqsp,
\end{align}
which we evaluate numerically. However, we find that this constraint is sub-dominant in all parts of parameter space, as it is always eclipsed by the bullet cluster constraint.

Finally, there also exist a constraint from observations of the \textit{Lyman-$\alpha$ forest}~\cite{Kolb:1990vq,Garzilli:2019qki} on the free-streaming length $\lambda_\text{fs}$, i.e.~$\lambda_\text{fs} < 0.24\,\mathrm{Mpc}$~\cite{Garzilli:2019qki}. However, we find that the resulting constraint is completely negligible for all relevant parameters, which is due to the fact that kinetic decoupling of $\phi$ happens rather late, thus leading to generally tiny values of the free-streaming length in our scenario. We will therefore not discuss these constraints any further.

\subsection{Domain walls}
\label{sec:domain_walls}
An interesting feature of spontaneously broken discrete symmetries, like $\mathbb{Z}_2$, is that they can lead to the formation of domain walls (DW). Such objects can be problematic if they are long-lived, as they might end up dominating the energy density of the Universe~\cite{Vilenkin:1984ib,Saikawa:2017hiv}. This, in turn, puts additional constraints on the scenario in question. Unlike DM candidates, whose energy density strongly depends on the DS temperature, the same is not true for DW. To see this, we follow~\cite{Saikawa:2017hiv} and consider a DW with surface tension $\sigma \sim \sqrt{\lambda}v^3$ and curvature radius $R_w$, which implies an energy density of the order $\rho'_w \sim \sigma/R_w$. While the radius of the walls grows with velocity $v_w$, i.e.~$R_w \simeq v_w t$, its motion in the HS bath at temperature $\Td$ is hampered by the friction due to particle pressure of the order $p'_w \sim v_w T'^4$ (we assume for simplicity relativistic particles). Assuming that the expansion of the Universe is driven by the visible sector, $t\sim M_\text{pl}/T$ and balancing $\rho'_w \sim p'_w$, we thus find $v_w \sim \xi^{-2} \sqrt{\sigma t}/M_\text{pl}$. Consequently, the wall velocity $v_w$ reaches the speed of light at $t_* \sim M_\text{pl}^2\xi^4/\sigma$. Hence, for $\xi < 1$, luminal velocities are reached earlier due to the reduced pressure. Now, since $R_w \propto t$, at $t=t_*$ the size of the DW is of the order of the size of the horizon, independent of $\xi$. From that point onward, it is argued and simulations show~\cite{Saikawa:2017hiv} that the DW form a self-similar network with an energy density scaling as $\rho'_w \sim \rho'_w(t_*)\times (t_*/t) \sim \sigma/t$, which is independent of $t_*$ and thus of $\xi$. Assuming that the Universe is dominated by radiation initially, it has been shown that such a network of DW starts to dominate the energy density at $t \sim 10^3\,\mathrm{s} \times (\mathrm{TeV}/\sigma)^3$. Taking matter-radiation equality, this implies that the surface tension of the DW is bounded from above by $\sqrt{\lambda} v^3 \sim \sigma \lesssim \mathrm{MeV}$~\cite{Zeldovich:1974uw,Saikawa:2017hiv}. In the left panel of fig.~\ref{fig:domain} we show the corresponding constraint in the $m_\phi-\xi_\text{fo}$ plane (purple, dashed line), noting that all points to the right of the line would be excluded. Note that the resulting limit implies a non-trivial relation between $m$ and $\xi$, since $\lambda$ is different for each point of parameter space, in order to produce the correct DM relic abundance.  

At face value, the DW constraints would rule out all the remaining parameter space for cannibal DM undergoing SSB. However, there exist certain loopholes for elevating these constraints. The simplest, yet not bullet-proof, option would be to assume that the symmetry was never restored in the HS, meaning that no domain wall formation happened in the first place. For this to work, the reheating temperature $\Td_\text{R}$ in the HS must be smaller than the critical temperature $\Td_c$ of the phase transition, i.e.~$\Td_\text{R}<\Td_c$. For cold dark sectors with $\xi \ll 1$, this is not unrealistic since $\Td_\text{R} \sim T_\text{R}\xi$  translates into a parametrically larger reheating temperature of the VS, i.e.~$T_\text{R} < \Td_c/\xi$, see fig.~\ref{fig:contours}. For instance, if $m_\phi \sim \mathrm{TeV}$, the corresponding reheating temperature should be less than $T_\text{R} \lesssim 10^3\,\mathrm{TeV}$ for $\xi \sim 10^{-3}$, which is compatible with the results in fig.~\ref{fig:domain}. However, following our scaling relations, lighter DM candidates require a larger value of $\xi$. Taking for instance $m_\phi \sim 10^{-1}\,\mathrm{MeV}$ and the corresponding value $\xi \sim 10^{-1}$, it is $T_R \sim \mathrm{MeV}$, which is borderline incompatible with standard cosmology~\cite{Hannestad:2004px}. An alternative loophole would be to assume that the DW are unstable and decay before they can dominate the energy density of the Universe~\cite{Vilenkin:1981zs}. For this to happen, the $\mathbb{Z}_2$ symmetry must be slightly broken, either explicitly within the HS or via some mixing with the VS. While the tiny amount of symmetry breaking that is required for this to happen does not affect our calculation of the freeze-out DM abundance, the DW decay can lead to an additional, non-thermal production of DM, whose contribution can even dominate the total DM abundance (cf.~e.g.~\cite{Hiramatsu:2012gg}). Moreover, decaying DW can lead to several new signatures and/or constraints, e.g.~by inducing DM decays into SM degrees of freedom or due to the emission of GW. Quantifying these effects is a non-trivial task, which is why we leave a corresponding discussion for future work.

\section{Conclusions}
\label{sec:conclusions}

Generating the DM abundance of the Universe within a cannibal DM scenarios is an old idea, which, however, has received renewed attention in recent years. Typically, such scenarios assume that the DM candidate is part of a HS, in which case it is natural to assume that its temperature is different from the one of the VS, i.e.~$\xi = T'/T \neq 1$. This enlarges the range of possible DM candidates, in particular towards heavier masses~\cite{Arcadi:2019oxh,Ghosh:2022asg}. In this work, we have expanded on this idea by classifying cannibal DM in the \textit{domain of thermal DM candidates}, while taking into account all relevant constraints, thus effectively mapping out the available parameter space for this class of scenarios.
Like a standard WIMP, cannibal DM can undergo non-relativistic freeze-out. However, due to entropy conservation in the HS, the resulting contours in parameter space instead share many similarities with those of relativistic relics, whose contours of constant relic abundance follow the relation $\xi \propto m_\text{DM}^{-\sfrac13 }$. In general, cannibal DM candidates thus lie between cold (i.e.~WIMPs) and hot relics within the aforementioned domain. 

To illustrate this point, we have focused on a real scalar DM particle with $\lambda \phi^4$ self-interactions. In the symmetric phase of this theory, DM stability is usually ensured via a discrete $\mathbb{Z}_2$ symmetry. This symmetry in turn can be either manifest (as already discussed in the literature) or spontaneously broken, with both possibilities considered in this work (cf.~sec.~\ref{sec:setup}). However, the case of SSB is especially interesting, since it features vanishing tree-level amplitudes for any process $k\phi \rightarrow \phi\phi$ with $k>2$ at threshold, a fact that is quite generic, and also holds true for $O(N)$ instead of $\mathbb{Z}_2$ symmetries. As a result, the tree-level cross-section governing DM annihilation in the broken phase is velocity suppressed around freeze-out, meaning that extra care is required in order to accurately determine the DM relic abundance. We have solved this problem numerically by employing a small set of physically motivated approximations, which enabled us to perform efficient calculations with $\sim$ 1\% accuracy (cf.~sec.~\ref{sec:evolution}). For comparison, we have also applied our formalism to the case without SSB and found very good agreement with the more intricate calculations of~\cite{Arcadi:2019oxh}, which solidifies our approach (see also app.~\ref{app:comparison}).

In addition, we have studied analytically the relation between the mass $m_\phi$ of the DM candidates and the temperature ratio $\xi_\text{fo}$ between the HS and VS at freeze-out (for fixed quartic coupling, cf.~\ref{sec:scaling}). In particular, we showed that cannibal DM candidates feature contours in the \textit{domain of thermal DM candidates} that are similar to the ones of relativistic relics, i.e.~$\xi \propto m_\phi^{-\sfrac13 (1 + \kappa)}$, but with an "anomalous scaling" $0 < \kappa \ll 1$ (cf.~fig.~\ref{fig:contours}). The so-obtained analytic results in fact turn out to be in excellent agreement with our numerical results (as well with the results of~\cite{Arcadi:2019oxh}). Interestingly, $\kappa \sim \mathcal{O}(x_\text{fo}'^{-1}) \ll 1$ and hence $\kappa \rightarrow 0$ in case the decoupling happens deep in the non-relativistic regime. In this case, the cannibal DM candidates feature the same scaling as hot relics, while still assuming a distinct relic abundance. 
In general, we found our results to be valid for a broad range of cannibal DM scenarios, meaning that they can shed some light on the peculiarities of such particles and how they compare to other DM abundance mechanisms.

As in the case of WIMPs, the domain of cannibal DM is additionally subject to various astrophysical and cosmological constraints, which we discuss explicitly (cf.~sec.~\ref{sec:domain}). The most relevant of these happens to be the unitarity bound on the cross-section of the dominant number-changing interaction, as it significantly reduces the allowed parameter space, especially compared to the usual WIMP scenario (cf.~fig.~\ref{fig:domain}). Additionally, SSB can lead to the formation of DW, which puts additional pressure on the parameter space. There do, however, exist certain loopholes to circumvent these constraints, i.e.~by allowing the DW to decay. To fully quantify these scenarios, i.e.~to calculate the amount of non-thermally produced DM (via DW decay) as well as the amount of emitted gravitational waves, is an interesting open question, which we leave for future work.

\section*{Acknowledgments}
We thank Simone Blasi, Rupert Coy, Camilo Garcia Cely and Alberto Mariotti for helpful discussions. This work is supported by the F.R.S./FNRS under the Excellence of Science (EoS) project No.~30820817 - be.h "The H boson gateway to physics beyond the Standard Model" and by the IISN convention No.~4.4503.15. The work of M.H.~is further supported by the F.R.S./FNRS.

\appendix

\section{The anomalous scaling}
\label{app:details}

In this appendix, we provide more details on how to obtain the relations for the anomalous scaling, specifically eqs.~\eqref{eq:scaling_main} and \eqref{eq:xcd_main}.

Given eq.~\eqref{eq:const_oh2}, contours of constant relic abundance in the non-relativistic limit are defined by the relation
\begin{align}
\text{d} \left[ m_\phi \left(x_\phi + \frac52 \right)^{-1} \xi_\infty^3 \right] \simeq 0\eqsp.
\end{align}
Defining $F(m_\phi, \xi_\infty) \equiv m_\phi \left[x_\phi(m_\phi, \xi_\infty) + \frac52\right]^{-1}\xi_\infty^3$ with $\text{d} F = 0$, we thus obtain
\begin{align}
0 = \left( \frac{\partial F}{\partial m_\phi} \right)_{\xi_\infty} \text{d}m_\phi + \left( \frac{\partial F}{\partial \xi_\infty} \right)_{m_\phi} \text{d}\xi_\infty\eqsp.
\end{align}
This can be rewritten as
\begin{align}
\frac{\text{d} \xi_\infty}{\text{d} m_\phi} = - \left( \frac{\partial F}{\partial m_\phi} \right)_{\xi_\infty} \Big/ \left( \frac{\partial F}{\partial \xi_\infty} \right)_{m_\phi}
\label{eq:dxi_dm}
\end{align}
with
\begin{align}
\frac{1}{F} \left( \frac{\partial F}{\partial m_\phi} \right)_{\xi_\infty} & = \frac{1}{m_\phi} - \frac{\partial_{m_\phi}x_\text{fo}'}{x_\text{fo}' + \frac52}\eqsp,\\
\frac{1}{F} \left( \frac{\partial F}{\partial \xi_\infty} \right)_{m_\phi} & = \frac{3}{\xi_\infty} - \frac{\partial_{\xi_\infty}x_\text{fo}'}{x_\text{fo}' + \frac52}\eqsp.
\end{align}
To determine the partial derivatives of $x_\text{fo}'$, we have to make use of an additional, independent equation. To this end, we utilize the fact that freeze-out happens approximately when
\begin{align}
\langle \sigma_k v^{k-1} \rangle n_\phi^{k-1} \sim H \sim \frac{T_\text{fo}^2}{M_\text{pl}}\eqsp.
\label{eq:rate_condition}
\end{align}
We then approximate the thermally averaged annihilation cross-section at freeze-out via the rather general expression
\begin{align}
\langle \sigma_k v^{k-1} \rangle \sim m_\phi^{-3k + 4} x_\text{fo}'^{-\beta}
\end{align}
with some value $\beta$, which covers the case of arbitrary $k$, as well as $s$-wave ($\beta = 0$) and $p$-wave etc. ($\beta \neq 0$) dominated processes. We then simplify eq.~\eqref{eq:rate_condition} by using $n_\phi \simeq s_\phi\left(x_\text{fo}'+\frac52\right)^{-1}$, $s_\phi = s_\text{VS}\times (s_\phi/s_\text{VS}) \sim \xi_\infty^3 T_\text{fo}^3$, as well as eq.~\eqref{eq:xicd} to translate $T_\text{fo} = \xi_\text{fo}^{-1} \Td_\text{fo} = m_\phi \xi_\text{fo}^{-1} x_\text{fo}^{-1}$ into a function of $\xi_\infty$. Overall, eq.~\eqref{eq:rate_condition} then yields
\begin{align}
\frac{3k-5}{3}\times x_\text{fo}' + &\left[ \frac{3k-5}{2} + \beta \right]\ln(x_\text{fo}') + \frac23 \ln(x_\text{fo}' + \frac52) \nonumber\\
& \simeq \text{const.} - \ln\left( \frac{m_\phi}{M_\text{pl}} \right) + 2\ln(\xi_\infty)\eqsp.
\label{eq:xfo}
\end{align}
This implies
\begin{align}
\frac{\partial x_\text{fo}'}{\partial \xi_\infty} \simeq \frac{2f_{k,\beta}(x_\text{fo}')}{\xi_\infty}\eqsp, \quad\frac{\partial x_\text{fo}'}{\partial \xi_\infty} \simeq -\frac{f_{k,\beta}(x_\text{fo}')}{m_\phi}\
\end{align}
with
\begin{align}
f_{k,\beta}(x_\text{fo}')^{-1} & = \frac{3k-5}{3} + \left(\frac{3k-5}{2} + \beta\right)\frac{1}{x_\text{fo}'} + \frac23 \frac{1}{x_\text{fo}'+\frac52}\eqsp.
\end{align}
After plugging this back into eq.~\eqref{eq:dxi_dm}, we thus obtain
\begin{align}
\frac{\text{d} \xi_\infty}{\text{d} m_\phi} \simeq -\frac{1}{3} \frac{\xi_\infty}{m_\phi} \frac{1+f_{k,\beta}(x_\text{fo}')/\left(x_\text{fo}' + \frac52\right)}{1-\frac23 f_{k,\beta}(x_\text{fo}')/\left(x_\text{fo}' + \frac52\right)}\eqsp.
\end{align}
Finally, this expression can be expanded in orders of $x_\text{fo}'^{-1}$, which yields
\begin{align}
\frac{\text{d}\xi_\infty}{\text{d} m_\phi} \simeq -\frac13 \frac{\xi_\infty}{m_\phi}\left[ 1 + \frac{5}{3k-5}\times \frac{1}{x_\text{fo}'} + \mathcal{O}\left(\frac{1}{x_\text{fo}'^2} \right) \right]\
\label{eq:dxi_dm_exp}
\end{align}
independent of $\beta$ (at first order).

\section{Comparison with previous results}
\label{app:comparison}

In this appendix, we compare our results for the case without SSB to some of the previous ones in the literature, specifically those that have been obtained in~\cite{Arcadi:2019oxh} and~\cite{Ghosh:2022asg}.

\begin{figure}[t]
\center
\includegraphics[width=0.6\textwidth]{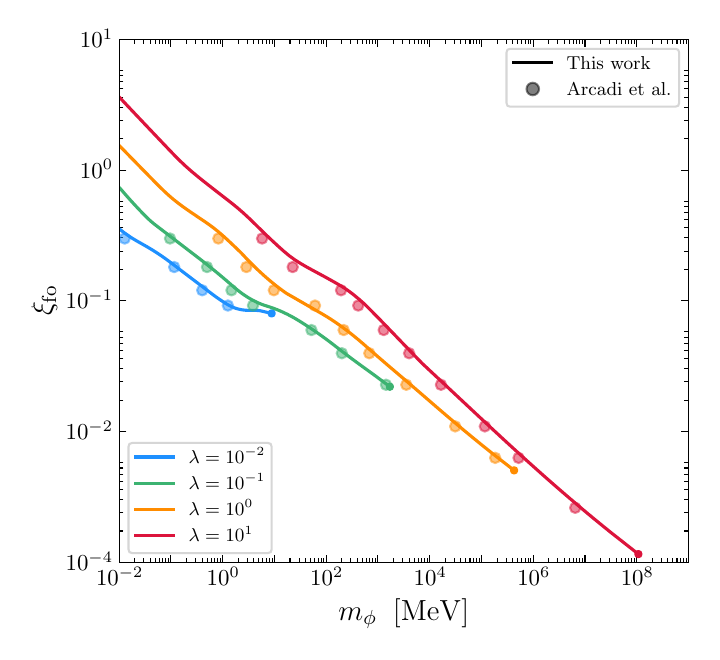}
\caption{Comparing of the results obtained in this paper (solid line) to those of~\cite{Arcadi:2019oxh} (transparent dots) in the case without SBB for different values of the quartic coupling $\lambda$ (different colours) in the $\xi_\text{fo}-m_\phi$ parameter plane.}
\label{fig:comp}
\end{figure}

A comparison of our results without SSB to those of~\cite{Arcadi:2019oxh} is indeed instructive, since -- in the aforementioned paper -- the evolution of the DM abundance is calculated with the full BE distribution even in the collision operator, contrary to our approximation in eq.~\eqref{eq:coll}. Consequently, the results of~\cite{Arcadi:2019oxh} are of high precision and hence a suitable baseline to judge our results against.
After digitizing the results of fig.~(8) in~\cite{Arcadi:2019oxh} and plotting them in the $m_\phi-\xi_\text{fo}$ plane, we obtain the results that are shown in fig.~\ref{fig:comp}.\footnote{Our results have been translated from the $m_\phi-\xi_\infty$ plane to the $m_\phi-\xi_\text{fo}$ plane by means of eq.~\eqref{eq:xicd}.} Here, solid lines represent the results obtained in this work, while the circles indicate the different data points that have been digitized from the reference paper.
Evidently, the results obtained in this work are in very good agreement with those in the reference paper. However, our results systematically lie above those of~\cite{Arcadi:2019oxh}. This is because, by assuming MB instead of BE statistics for $\mathcal{C}_{\phi\phi\rightarrow k\phi}$ in eq.~\eqref{eq:cfull}, we slightly overestimate the collision operator. Consequently, the particle is kept in equilibrium a little too long, meaning that too large values of $\xi_\infty$ and thus $\xi_\text{fo}$ are needed to obtain the correct relic abundance. Nevertheless, these deviations are sufficiently small, and we find a general agreement at the $\mathcal{O}(1\%)$ level. Especially, the results in~~\cite{Arcadi:2019oxh} thus also follow and consequently validate our general scaling relation in eq.~\eqref{eq:scaling_main}.
Overall, this comparison therefore indicates that the method used in this paper -- albeit less sophisticated than the one in~\cite{Arcadi:2019oxh} -- is suitable to accurately describe the decoupling process of cannibal DM. It is thus sufficient to utilize the simpler mechanism presented in this paper (cf.~sec.~\ref{sec:evolution}), which greatly reduces the computational complexity.

Additionally, our general findings can also be compared to the results obtained in~\cite{Ghosh:2022asg}. In this paper, a general cubic coupling is assumed in the Lagrangian, similar to eq.~\eqref{eq:lag_gen}, but their chosen parameter points never feature a vanishing of the $s$-wave contribution to the annihilation cross-section. Nevertheless, the general findings of this paper should still be comparable with our scaling relation in eq.~\eqref{eq:scaling_main}.
After digitizing the black lines of fig.~(3) in~\cite{Ghosh:2022asg},\footnote{Note that the definition of the temperature ratio in~\cite{Ghosh:2022asg} is inverse to the one used in this paper.} we find, however, that the obtained contours do not follow the general scaling relations derived in this work. Specifically, the lines in~\cite{Ghosh:2022asg} do not increase monotonously for smaller masses.  While there is no immediate explanation for this discrepancy, we reiterate that our numerical results agree with our analytical results as well as with the numerical ones presented in~\cite{Arcadi:2019oxh}.

\bibliographystyle{JHEP}
\bibliography{refs}

\end{document}